\documentclass[a4paper]{spie}  
\usepackage[]{graphicx}
\usepackage{amssymb,amstext}

\def\sun{\hbox{$\odot$}}
\def\arcmin{\hbox{$^\prime$}}
\def\arcsec{\hbox{$^{\prime\prime}$}}

\def\farcs{\hbox{$.\!\!^{\prime\prime}$}}

\def\mag{\ifmmode^{\rm m }\else$^{\rm m}$\fi}
\def\fracarcsec{\hbox{$.\!\!^{\rm s}$}}

\def\as{$\,^{\prime\prime}\,$}

\def\hh{\ifmmode^{\rm h}\else$^{\rm h}$\fi}
\def\mm{\ifmmode^{\rm m}\else$^{\rm m}$\fi}
\def\ss{\ifmmode^{\rm s}\else$^{\rm s}$\fi}
\def\deg{\ifmmode^\circ\else$^\circ $\fi}
\def\amin{\ifmmode^\prime\else$^\prime $\fi}
\def\rahm#1#2{\ifmmode{#1}\else{$#1$}\fi\hh\ifmmode{#2}\else{$#2$}\fi\mm}
\def\decdm#1#2{\ifmmode{#1}\else{$#1$}\fi\deg\ifmmode{#2}\else{$#2$}\fi\amin}
\def\ras#1#2{\ifmmode{#1}\else{$#1$}\fi\fracarcsec\ifmmode{#2}\else{$#2$}\fi}
\def\decs#1#2{\ifmmode{#1}\else{$#1$}\fi\farcs\ifmmode{#2}\else{$#2$}\fi}
\def\ra#1#2#3{\ifmmode{#1}\else{$#1$}\fi\hh\ \ifmmode{#2}\else{$#2$}\fi\mm\ \ifmmode{#3}\else{$#3$}\fi\ss}
\def\dec#1#2#3{\ifmmode{#1}\else{$#1$}\fi\deg\ \ifmmode{#2}\else{$#2$}\fi\amin\ \ifmmode{#3}\else{$#3$}\fi\as}

\newcommand{\rab}[4]{\ifmmode{#1}\else{$#1$}\fi\hh\ \ifmmode{#1}\else{$#2$}\fi\mm\ \ifmmode{#1}\else{$#3$}\fi\fracarcsec\ifmmode{#1}\else{$#4$}\fi}
\newcommand{\decb}[4]{\ifmmode{#1}\else{$#1$}\fi\deg\ \ifmmode{#1}\else{$#2$}\fi\amin\ \ifmmode{#1}\else{$#3$}\fi\farcs\ifmmode{#1}\else{$#4$}\fi}

\title{First results from VLTI near-infrared interferometry\\ on high-mass young stellar objects} 

\author{Stefan Kraus\supit{a}, Karl-Heinz Hofmann\supit{b}, Karl M. Menten\supit{b}, Dieter Schertl\supit{b}, Gerd Weigelt\supit{b}, Friedrich Wyrowski\supit{b}, Anthony Meilland\supit{b,c}, Karine Perraut\supit{d}, Romain Petrov\supit{c}, \\
Sylvie Robbe-Dubois\supit{c}, Peter Schilke\supit{e} and Leonardo Testi\supit{f}
\skiplinehalf
\supit{a}Department of Astronomy, University of Michigan, 500 Church St., Ann Arbor, MI 48103, USA; 
\supit{b}Max-Planck-Institut f\"{u}r Radioastronomie, Auf dem H\"{u}gel 69, 53123 Bonn, Germany; \\
\supit{c}Laboratoire Hippolyte Fizeau, UMR 6525 Universit\'{e} de Nice Sophia-Antipolis/CNRS/Observatoire de la C\^{o}te d'\,Azur, Parc Valrose, 06108 Nice Cedex 2, France; 
\supit{d}Laboratoire d'Astrophysique de Grenoble, UMR 5571 Universit\'{e} Joseph Fourier/CNRS, BP 53, 38041 Grenoble Cedex 9, France; \\
\supit{e}I.~Physikalisches Institut, Universit\"{a}t zu K\"{o}ln, Z\"{u}lpicher Strasse 77, 50937 Cologne, Germany; \\
\supit{f}INAF-Osservatorio Astrofisico di Arcetri, Largo Fermi 5, 50125 Firenze, Italy
}

\authorinfo{Further author information: (Send correspondence to S.K.)\\
E-mail: stefankr@umich.edu, Telephone: 1 734 615 7374}

\begin{document} 
  \maketitle 

\begin{abstract}
Due to the recent dramatic technological advances,
infrared interferometry can now be applied to new classes of objects, 
resulting in exciting new science prospects,
for instance, in the area of high-mass star formation.
Although extensively studied at various wavelengths, the process 
through which massive stars form is still only poorly understood.  
For instance, it has been proposed that massive stars might form
like low-mass stars by mass accretion through a circumstellar
disk/envelope, or otherwise by coalescence in a dense stellar cluster.
Therefore, clear observational evidence, such as the 
detection of disks around high-mass young stellar objects (YSOs), 
is urgently needed in order to unambiguously identify the formation mode of the most massive stars.

After discussing the technological challenges which result from the
special properties of these objects, we present first near-infrared 
interferometric observations, which we obtained on the massive YSO IRAS\,13481-6124
using VLTI/AMBER infrared long-baseline interferometry and NTT speckle interferometry.
From our extensive data set, we reconstruct a model-independent aperture synthesis image
which shows an elongated structure with a size of $\sim 13\times19$~AU, 
consistent with a disk seen under an inclination of $\sim45^{\circ}$.
The measured wavelength-dependent visibilities and
closure phases allow us to derive the radial disk temperature gradient and 
to detect a dust-free region inside of 9.5~AU from the star, revealing qualitative 
and quantitative similarities with the disks observed in low-mass star formation. 
In complementary mid-infrared Spitzer and sub-millimeter APEX imaging observations
we detect two bow shocks and a molecular outflow, 
which are oriented perpendicular to the disk plane and 
indicate the presence of a bipolar outflow emanating from the inner 
regions of the system.
\end{abstract}



\section{INTRODUCTION}
\label{sec:intro}  

In recent years, the field of infrared interferometry has experienced 
dramatic technological progress, providing substantial improvements, for instance, 
in terms of the available baseline lengths, spectral resolution, and 
spectral coverage.
Besides quantitative improvements, the instrumentational 
and infrastructure advances have also considerably extended the accessible 
range of astrophysical targets, enabling infrared interferometry
to address key scientific areas such as high-mass star formation.

Although scarce in number, massive stars ($> 10$~M$_{\sun}$) are of
fundamental importance for astrophysics.
Already shortly after their birth, they start to disperse their natal
molecular clouds with strong stellar outflows and photoevaporate the
protoplanetary disks around nearby low-mass stars.  
After a short hydrogen-burning phase, the most massive stars end their
evolution as supernovae, enriching the interstellar medium with heavy elements
and triggering the next generation of star formation through supernova shock
waves.  In spite of their significance, it is still poorly understood how
high-mass stars form.

Early spherically symmetric calculations suggested that the accretion scenario, which appears well established
for stars with masses below 10~M$_{\sun}$, might not work for massive protostars due to the strong radiation {pressure\cite{kah74,wol87}}. 
Therefore, it was proposed that high-mass stars might not form by accretion, but instead by stellar {merging\cite{bon98}}. 
Although recent theoretical work showed that the radiation pressure limit might be overcome when considering more 
complex infall geometries than spherically symmetric {ones\cite{mck02,kru09}}, clear observational evidence, 
such as the detection of compact dusty disks around massive young stellar objects, 
is needed to unambiguously identify the formation mode of the most massive stars.

Interferometric observations at radio wavelengths already provided fascinating insight into 
the distribution and kinematics of the circumstellar material, although the existing observations 
are still strongly limited by the achievable angular resolution of thousands to hundreds of astronomical units (AUs). 
For instance, millimeter {observations\cite{bel06b}} with an angular resolution of {$\sim 0.8$\arcsec} 
revealed a large ($\sim 10^{4}$~AU) rotating toroidal structure and gas infall motion around the YSO G24.78+0.08.
Recently, mid-infrared long-baseline interferometric {observations\cite{lin09,dew09}} 
probed the environment around two massive YSOs on scales of a few tens of AUs
and could resolve the inner regions of an envelope around these stars.
In order to separate the contributions from a circumstellar disk
and to avoid confusion with the surrounding rotating circumstellar envelope,
observations in the near-infrared around $\sim 2~\mu$m are very promising,
since this wavelength regime is sensitive to the
thermal emission of hot ($> 1000$~K) gas and dust,
probing temperatures where dust is expected to
sublimate ($T \sim 1500$...2000~K) and to form a dust-free inner hole. 
By forming telescope baselines up to hundreds of meters, infrared interferometric imaging can now 
provide the milli-arcsecond (mas) angular resolution that is required to directly 
resolve these inner disk regions. Compared to earlier studies on high-mass YSOs, which used 
conventional imaging techniques at 10\,m class telescopes equipped with adaptive optics systems, 
the gain in resolving power is at least one order of magnitude, 
while the gain compared to state-of-the-art (sub-)millimeter disk studies is about two orders of magnitudes. 

However, the technological challenges for infrared interferometry on young high-mass stars are high.
First, massive YSOs are still deeply embedded in their natal clouds, which results in an extremely steep 
drop of the spectral energy distribution (SED) between mid-infrared and near-infrared wavelengths.
Accordingly, most massive YSOs are not associated with a visual counterpart.
Therefore, it is essential to use off-axis telescope guiding, as offered by the STRAP units at
VLTI 1.8\,m auxiliary telescopes (ATs), which use Avalanche Photodiodes 
to provide a tip-tilt correction for guide stars brighter than $V=13.5$.
The VLTI 8.2\,m unit telescopes (UTs) are equipped with MACAO adaptive optic units
operating on natural guide stars down to $V=17$.
These sensitivity limits and the requirement that the guide star has to be located within the isoplanatic patch ($\sim${1\arcmin}), 
makes it difficult to find suitable guide stars for many high-mass YSOs, 
in particular since the number of potential off-axis guide stars is also
strongly reduced by the high visual extinction in high-mass star forming regions.

Recently, we have obtained first near-infrared interferometric observations 
of a young high-mass {star\cite{kra10}}, namely IRAS\,13481-6124. 
This star has a nearby foreground star ($V=12.6$) at a separation of {$\sim 17$\arcsec}, which can be used for telescope guiding even with the ATs. 
Furthermore, IRAS\,13481-6124 is located at a very favourable celestial location (DEC=$-61^{\circ}$)
which results in rather circular $uv$-tracks, providing optimal prerequisits for earth-rotation aperture-synthesis imaging.
Earlier studies have estimated that IRAS\,13481-6124 is located at a distance of about 3.5~kpc\cite{fon05} and harbours a central object with a mass of 
{$\sim 20$~M$_{\sun}$\cite{gra09}}, which is embedded in a cloud with a total mass of {$\sim 1500$~M$_{\sun}$\cite{bel06}}.

\newpage

\section{OBSERVATIONS} 
\label{sec:observations}

\begin{figure}[t]
  \centering
  $\begin{array}{c@{\hspace{5mm}}c}
    \includegraphics[height=6.6cm,angle=0]{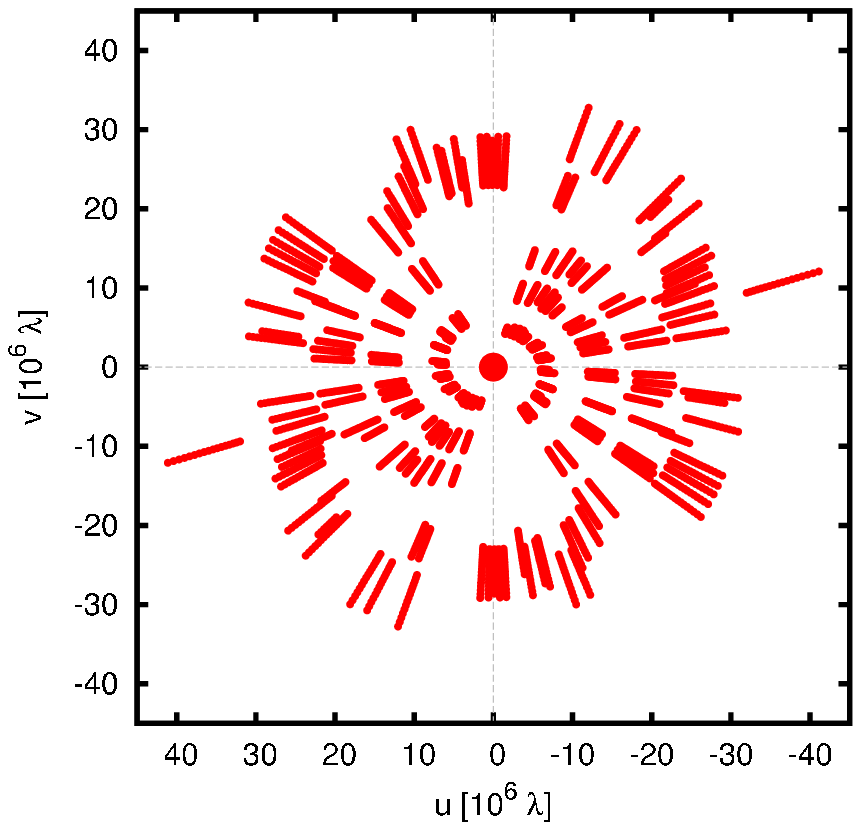} & \includegraphics[height=6.6cm]{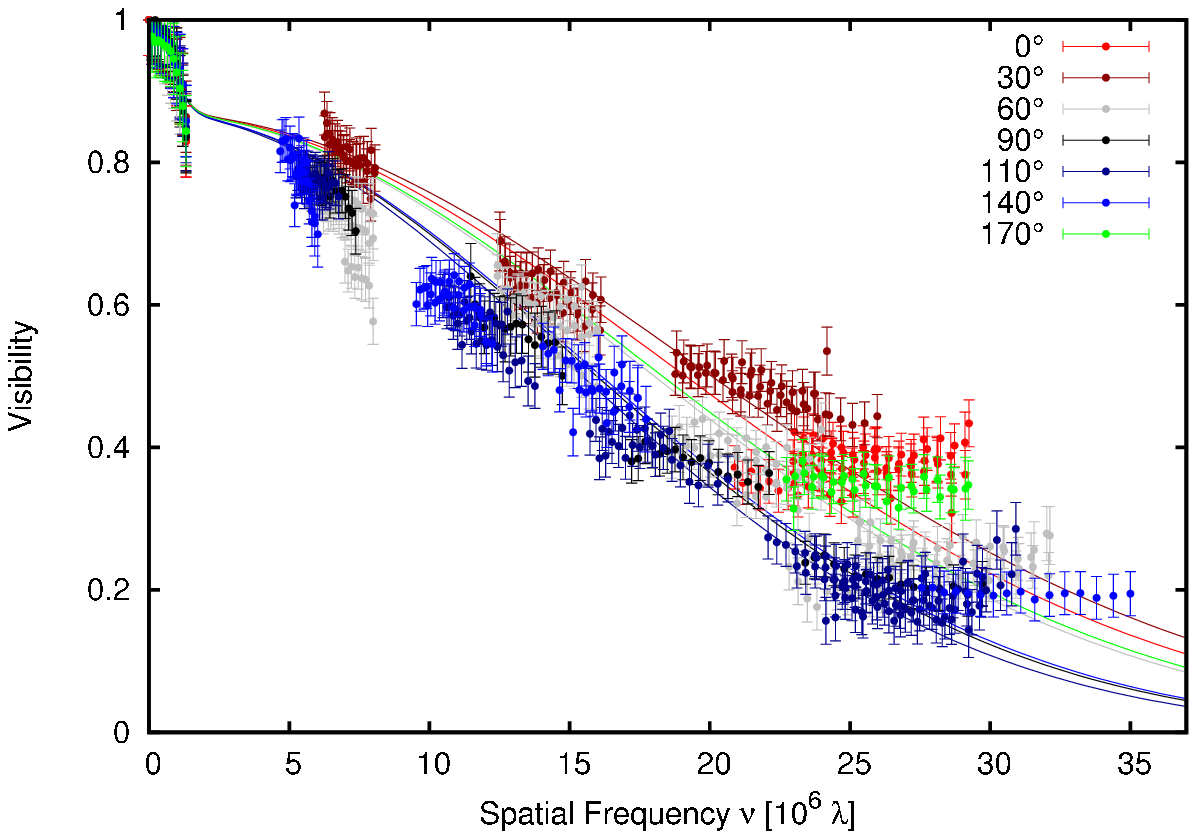} \\
  \end{array}$
  \caption{{\it Left:} $uv$-plane coverage obtained with our 
    VLTI/AMBER and NTT/speckle observations.
    {\it Right:} Visibilities recorded with our VLTI/AMBER and NTT/speckle
    observations towards different position angles, revealing a strong
    position angle-dependence of the visibility function.
  }
  \label{fig:uvcov}
\end{figure}

\begin{figure}[p]
  \centering
  \includegraphics[width=16cm]{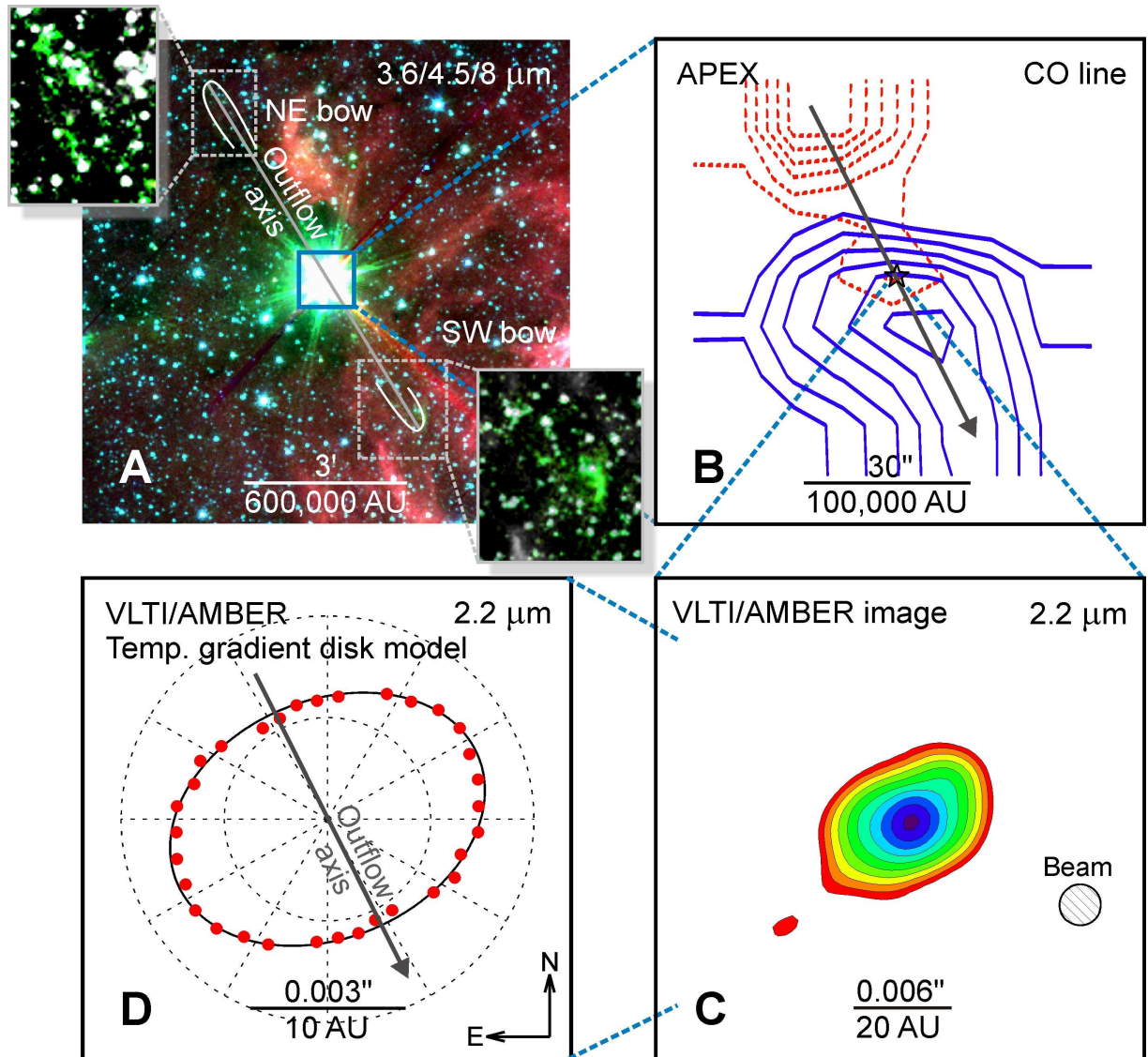} 
  \caption{
    Zoom in on IRAS\,13481-6124, covering spatial scales over more than five orders of magnitudes:
    {\it (A)}~In Spitzer/IRAC images, we detect two bow-shock structures, indicating
    a collimated outflow. 
    {\it (B)}~The outflow is also detected in molecular line emission using APEX/SHFI on scales
    of a few 10\,000 AU (the contours show blue-shifted velocities of --10 km\,s$^{-1}$ and redshifted velocities of +5 km\,s$^{-1}$,
    where the latter are indicated with dashed lines).
    {\it (C)}~Applying our image reconstruction algorithm to the VLTI/AMBER data, we reconstructed a
    model-independent aperture synthesis image of IRAS\,13481-6124. The image reveals an elongated structure,
    which is oriented perpendicular to the outflow direction (contours decrease from peak intensity by factors of $\sqrt{2}$). 
    {\it (D)}~Elongation of the compact emission component, as determined with our temperature gradient disk model.
  }
  \label{fig:overview}
\end{figure}

We observed IRAS\,13481-6124 during seven nights in 2008 and 2009
using the near-infrared three-telescope beam combination instrument AMBER \cite{pet07}
at ESO's Very Large Telescope Interferometer (VLTI).
Our spectro-interferometric observations cover the near-infrared $K$-band 
between $\sim 1.95$ to $2.55~\mu$m with a spectral resolution $\lambda/\Delta\lambda=35$.
Using the VLTI/ATs on
three different array configurations (E0-G0-H0, D0-H0-G1, A0-D0-H0)
and at various hour angles, we could achieve a 
relatively good $uv$-coverage (Fig.~\ref{fig:uvcov}, {\it left}), including baseline lengths between 12 and 85\,m.
In order to cover spatially extended structures, we complemented 
the VLTI long-baseline interferometric observations with 
bispectrum speckle interferometric {observations\cite{lab70,wei77}},
yielding precise visibility information for baselines $\lesssim 3.5$\,m.
The speckle $K$-band data was recorded with ESO's New Technology Telescope (NTT)
and our visitor camera, employing a Rockwell HAWAII detector.
Following the standard observing procedure, both the VLTI and 
speckle observations on IRAS\,13481-6124 were interlayed 
with calibrator observations in order to derive absolute 
calibrated visibilities and closure phases, which we then
used for our image reconstruction and model fitting approach.

Besides these high-resolution $K$-band continuum observations, which
directly trace the thermal dust emission and scattered light in
the circumstellar environment around IRAS\,13481-6124, 
we searched for signatures of mass outflow on larger spatial scales.
For this, we obtained archival {\it Spitzer}/IRAC images
which were recorded in the course of the GLIMPSE survey\cite{ben03} and
which cover wavelengths around $3.6$, $4.5$, $5.8$, and $8.0~\mu$m.
In the IRAC $4.5~\mu$m image (codes as green color in the
color composite shown in Fig.~\ref{fig:overview}A)
we detect two bow shock structures which are oriented
along position angle $31\pm 6^{\circ}$ and likely indicate 
H$_{2}$ shock-tracer line {emission\cite{smi06}}.
In order to trace a potential molecular outflow, we obtained 
a $^{12}$CO\,(3-2, 346~GHz) map using the SHFI heterodyne receiver 
at the APEX~12\,m telescope located on the Chajnantor plateau in the Chilean Andes.
The 
In the obtained CO channel maps, we detect a bipolar outflow
oriented along position angle $26\pm 9^{\circ}$, 
with the approaching (blue-shifted) lobe southwest of IRAS\,13481-6124
(Fig.~\ref{fig:overview}B).

\section{APERTURE SYNTHESIS IMAGING}

Given that our observations provide a good $uv$-coverage and include
closure phase information, we could use our data
to reconstruct a model-independent aperture synthesis image.
For this purpose, we employed the Building Block Mapping
(BBM) image reconstruction {algorithm\cite{hof93}}.
This algorithm has already been employed successfully
in various earlier projects to reconstruct 
interferometric images from simulated\cite{law06,cot08}
and long-baseline interferometric {data\cite{kra07,kra09a,mil09}}.
The reconstruction was performed on a $256\times256$ grid
using a pixel scale of $0.85$~mas/pixel and convergence was 
reached after $\sim 12\,000$ iteration steps.
In order to maximize the $uv$-coverage for imaging, 
we combined data from all spectral channels in order to
reconstruct one monochromatic image.
The obtained image was convolved with a Gaussian of
FWHM 2.4~mas, corresponding to a resolution of 
$\lambda/2B$, where $\lambda$ denotes the observing wavelength
and $B$ the length of the longest employed baseline.

The reconstructed image (Fig.~\ref{fig:overview}C) 
has a reduced goodness-of-fit value of 1.5 and 1.0 
for the squared visibilities and the closure phases, respectively.
The residuals are well distributed, i.e.\ they closely
follow a normal distribution.
The image clearly reveals an elongated structure which is oriented in
north-western direction. Besides this compact component, some flux appears distributed 
over the whole image.
Measuring the flux in different apertures, we find that 
80\% of the total image flux is contained in a $7.0$~mas aperture,
while the remaining flux elements are spread rather homogeneously over the image.

\section{MODELING}
\label{sec:modeling}

In order to further characterize the elongated structure detected with
aperture synthesis imaging, we fitted geometric and detailed physical models to our data, 
allowing us to deduce object information on smaller scales than the diffraction-limited resolution 
of our reconstructed image.

\subsection{Geometric models}
\label{sec:geommodeling}

\begin{figure}[p]
  \centering
  \includegraphics[width=7cm]{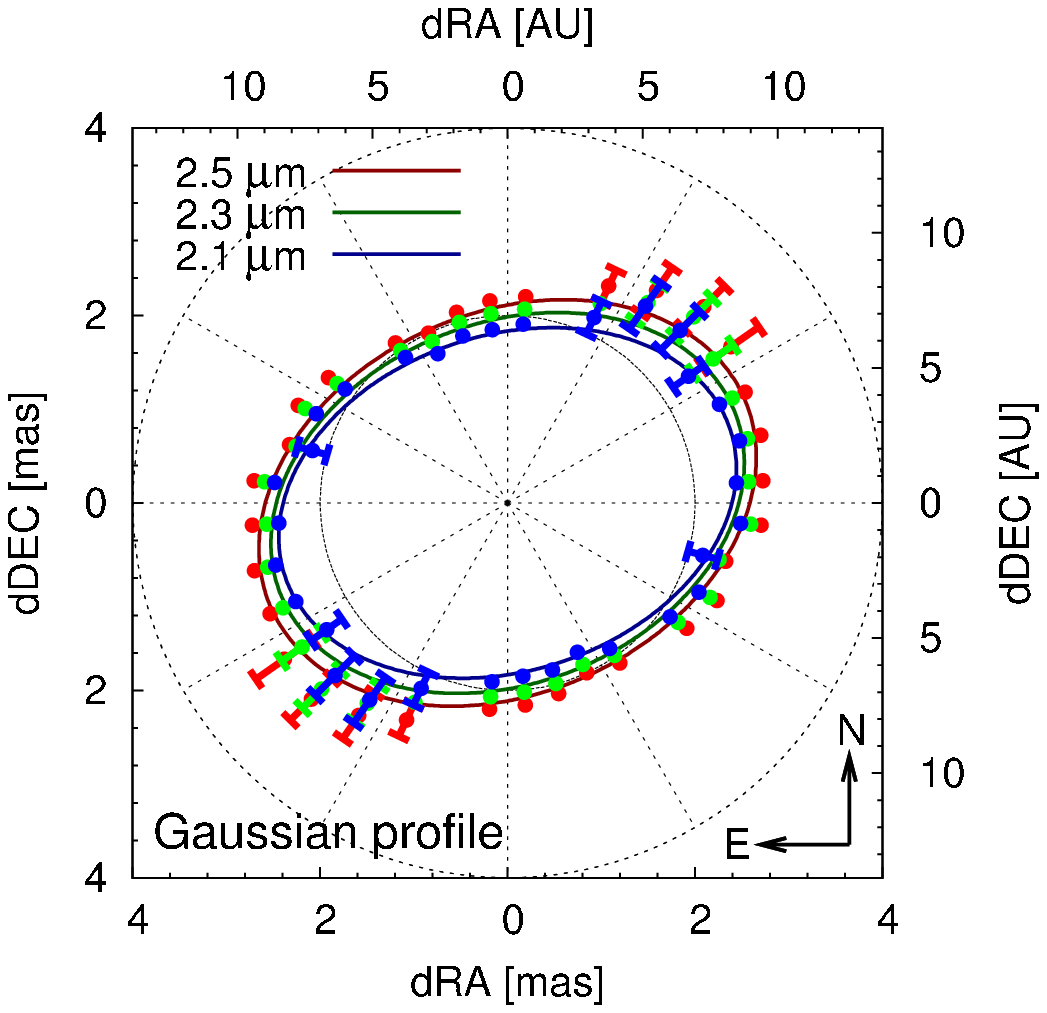}\\
  \includegraphics[width=13.5cm]{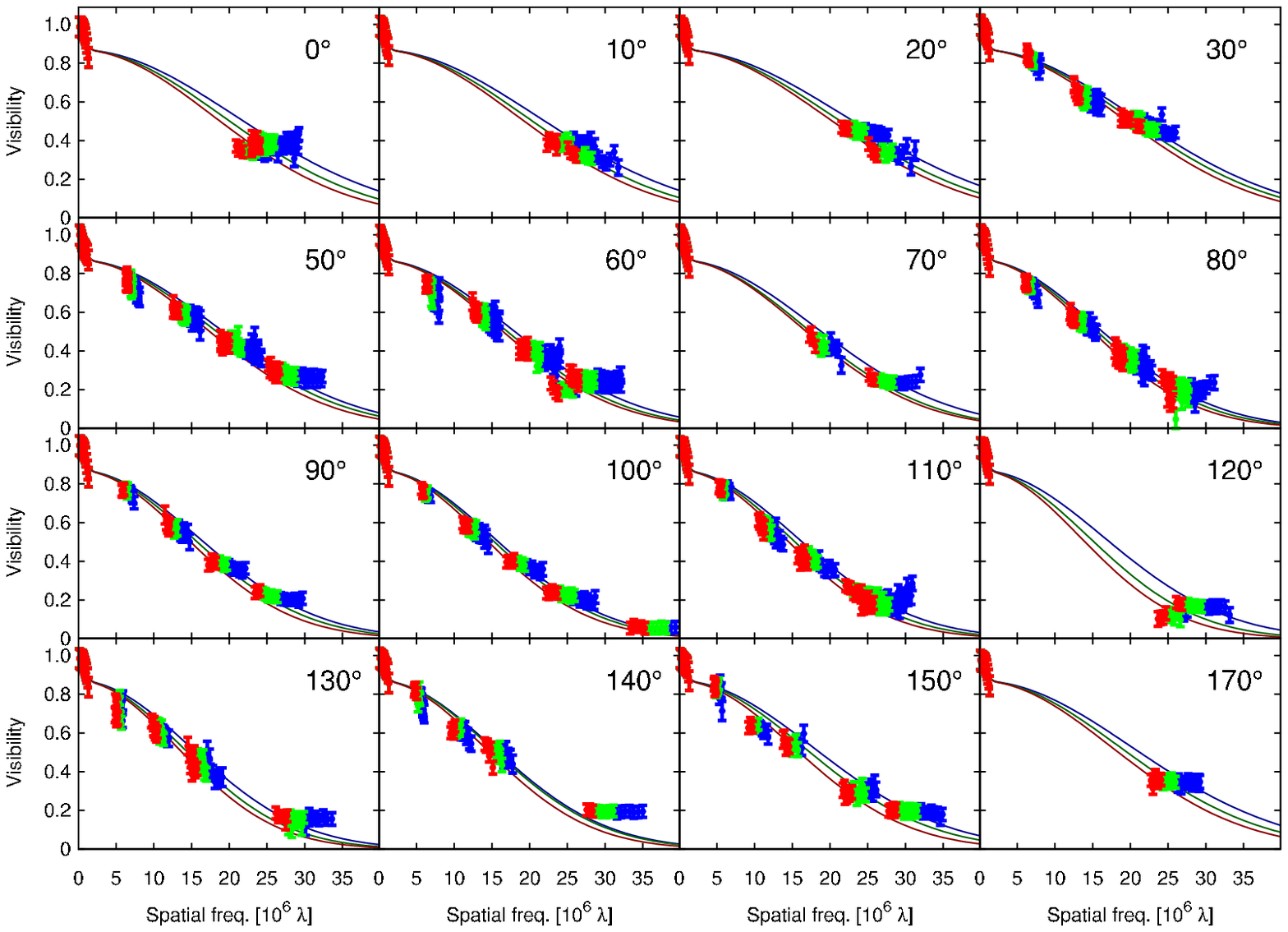}
  \caption{Elongation of the near-infrared emitting region {\it (top)},
    as determined by fitting Gaussian models to
    independent subsets of our visibility data {\it (bottom)}, 
    each covering a position angle range of $10${\deg}.
    In order to investigate whether the object morphology
    changes with wavelength, we have 
    divided our data in three wavelength bins
    (blue: $\lambda < 2.2$; 
    green: $2.2 \leq \lambda < 2.4$; 
    red: $\lambda \geq 2.4$) and then adjusted the
    Gaussian FHWM diameter.
    Besides the compact component, this model also includes a Gaussian with FWHM diameter 108~mas, 
    representing the extended envelope component detected with our NTT/speckle observations.
  }
  \label{fig:elonggauss}
\end{figure}

Our measurements show that the visibility function drops rapidly at baseline lengths $\lesssim 3$~m, 
but stays then nearly constant up to baseline lengths of about 12~m, followed by a uniform, 
nearly linear decline at baseline lengths up to 85~m (Fig.~\ref{fig:uvcov}, {\it right}), 
providing clear evidence for the presence of at least two spatial components. 
We use a Gaussian model to estimate the characteristic size  of the two components and 
find that the first, extended component has a FWHM diameter of about 108~mas and contributes $\sim15$\% of the total flux. 
Based on the finding of scattered light envelopes around other massive stars, we suggest that this extended, 
centro-symmetric component, which also appears in our image as homogenously distributed flux elements, 
is the scattered light contribution of a natal envelope. 
The second, compact component accounts for 85\% of the total $K$-band flux and 
has a size of $5.4\times3.8$~mas, consistent with a disk seen under an inclination angle of $45^{\circ}$. 
Furthermore, we find that the size of the compact component increases significantly towards longer wavelengths, 
as shown by the measured equivalent widths in Fig.~\ref{fig:elonggauss} ({\it top}). 
This effect, which was found towards many low- and intermediate-mass YSO {disks\cite{eis07a,kra08a}}, 
is likely indicating a temperature gradient in the circumstellar material, where hotter material 
(radiating more effectively at shorter wavelengths) is located closer to the star.

Motivated by these indications for an internal temperature structure, we fitted analytical disk models 
with a radial temperature power-law, $T(r)=T_{\rm in} (r/r_{\rm in})^{-q}$, to our data. Assuming reasonable values for 
the inner disk temperature, i.e. $T_{\rm in}=1500$ to 2000~K, we find that this model can reproduce our 
interferometric data with an inner disk radius $r_{\rm in}$ of 3~mas (9.5 AU), and a temperature power law index 
$q$ of 0.4, which is consistent with the theoretical temperature gradient of flared irradiated disks\cite{chi97} ($q=0.43$). 
Compared to other geometric models (e.g.\ uniform disk model with $\chi^2=4.48$ or Gaussian model with $\chi^2=2.56$),
the temperature gradient disk model provides a significantly better representation of our data ($\chi^2=1.40$).
Intriguingly, the derived inner disk radius of 9.5~AU agrees with the expected location where dust in 
an irradiated circumstellar disk would sublimate, i.e.\ 6.2~AU to 10.9~AU 
assuming grey dust and dust sublimation temperatures between 2000 and 1500~K. 
Therefore, IRAS\,13481-6124 is following the size-luminosity {relation\cite{mon05}}, 
which is well established for low- to intermediate mass YSOs, suggesting that the near-infrared emission 
mainly traces material at the dust sublimation radius, similar as in the disks around T~Tauri and Herbig~Ae/Be stars.

The derived disk orientation ($120^{\circ}$, Fig.~\ref{fig:overview}D) is perpendicular to the determined outflow axis 
($26\pm 9^{\circ}$ and $31\pm 6^{\circ}$), 
suggesting that the AU-scale disk resolved by VLTI/AMBER 
is indeed the driving engine of the detected collimated outflow.

\subsection{Radiative transfer modeling}
\label{sec:RTmodeling}

\begin{figure}[p]
  \centering
      $\begin{array}{c@{\hspace{5mm}}c}
        \includegraphics[width=7.7cm,angle=0]{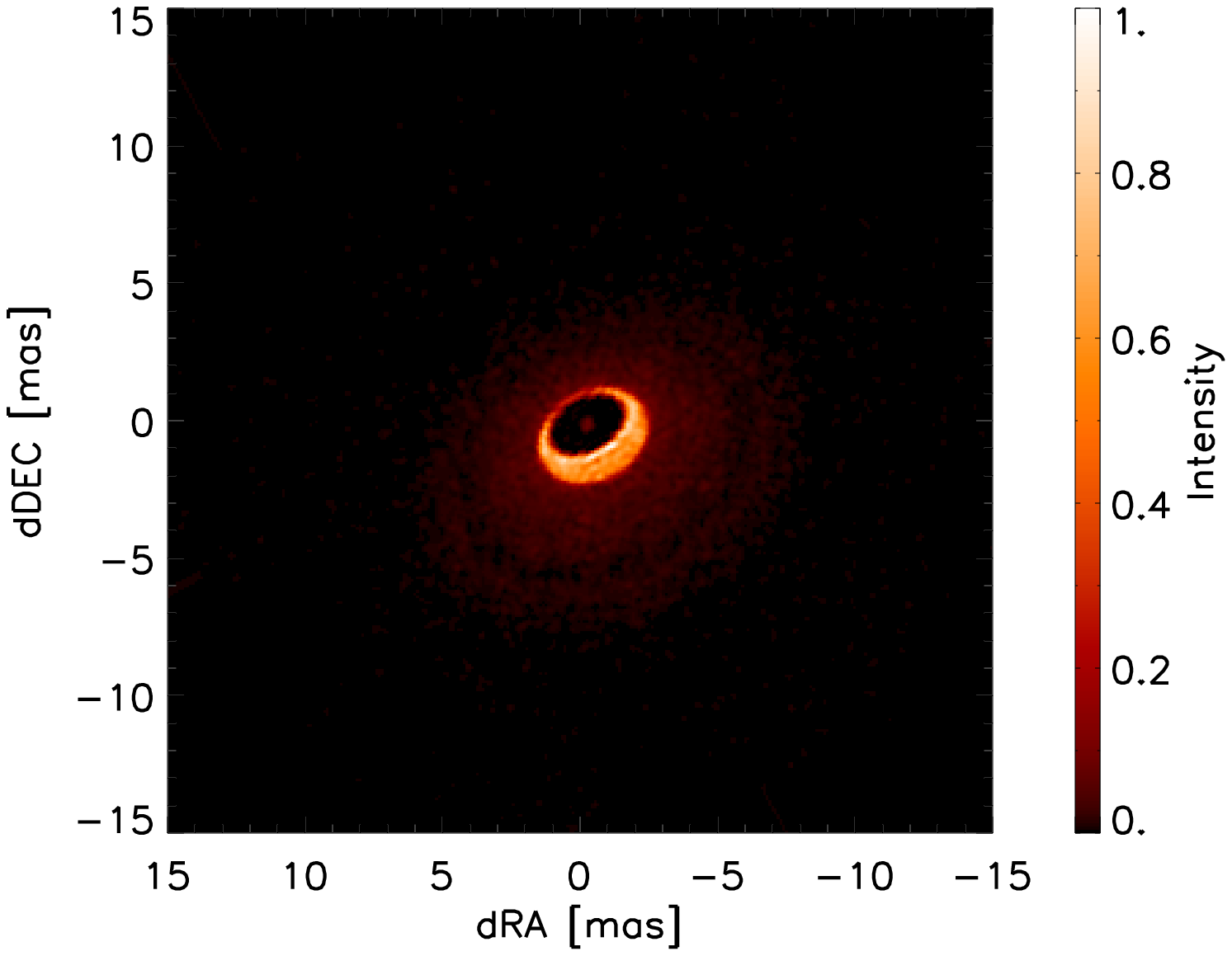} & \includegraphics[width=8.2cm]{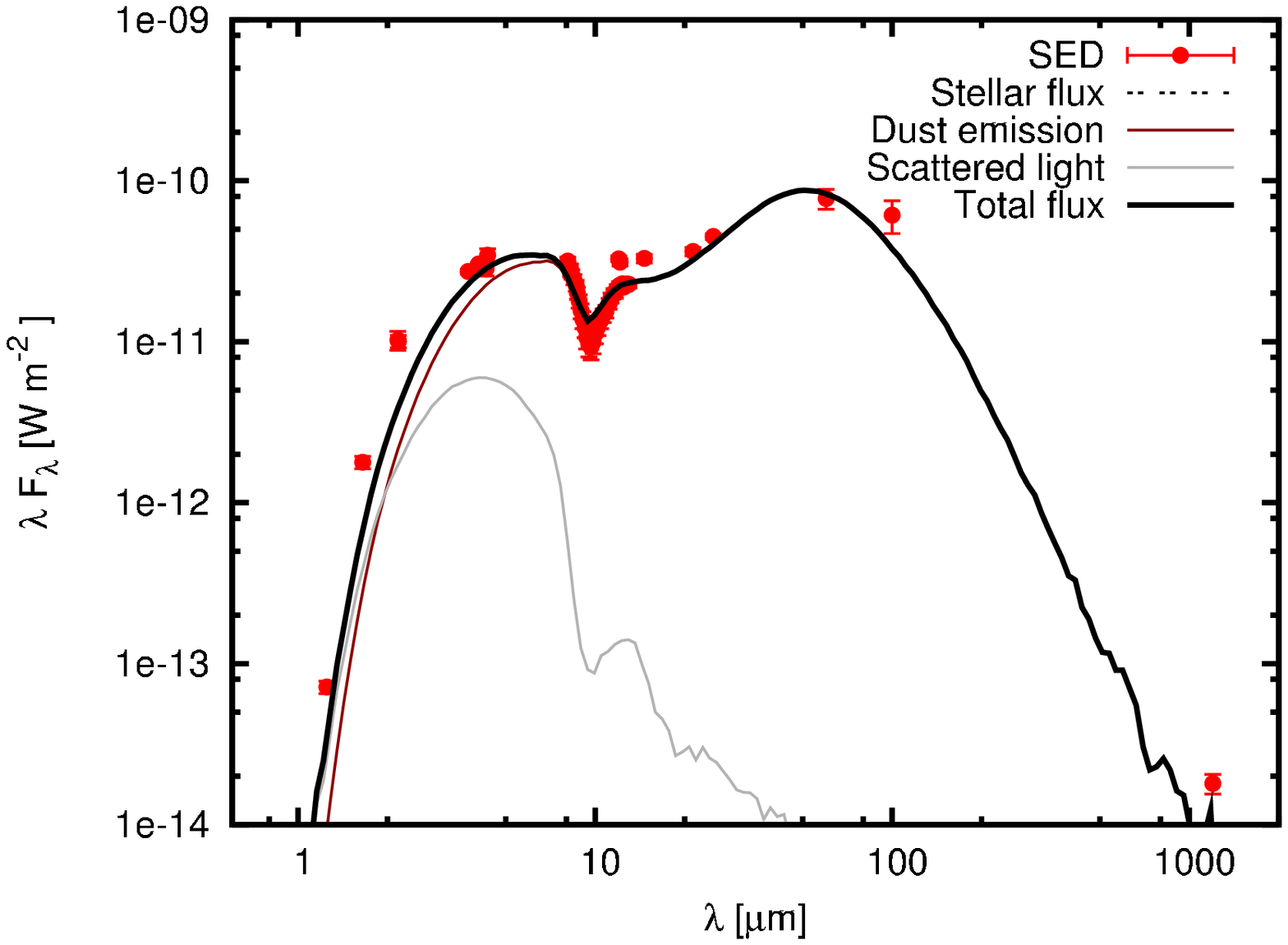} \\
      \end{array}$
  \includegraphics[width=16.5cm]{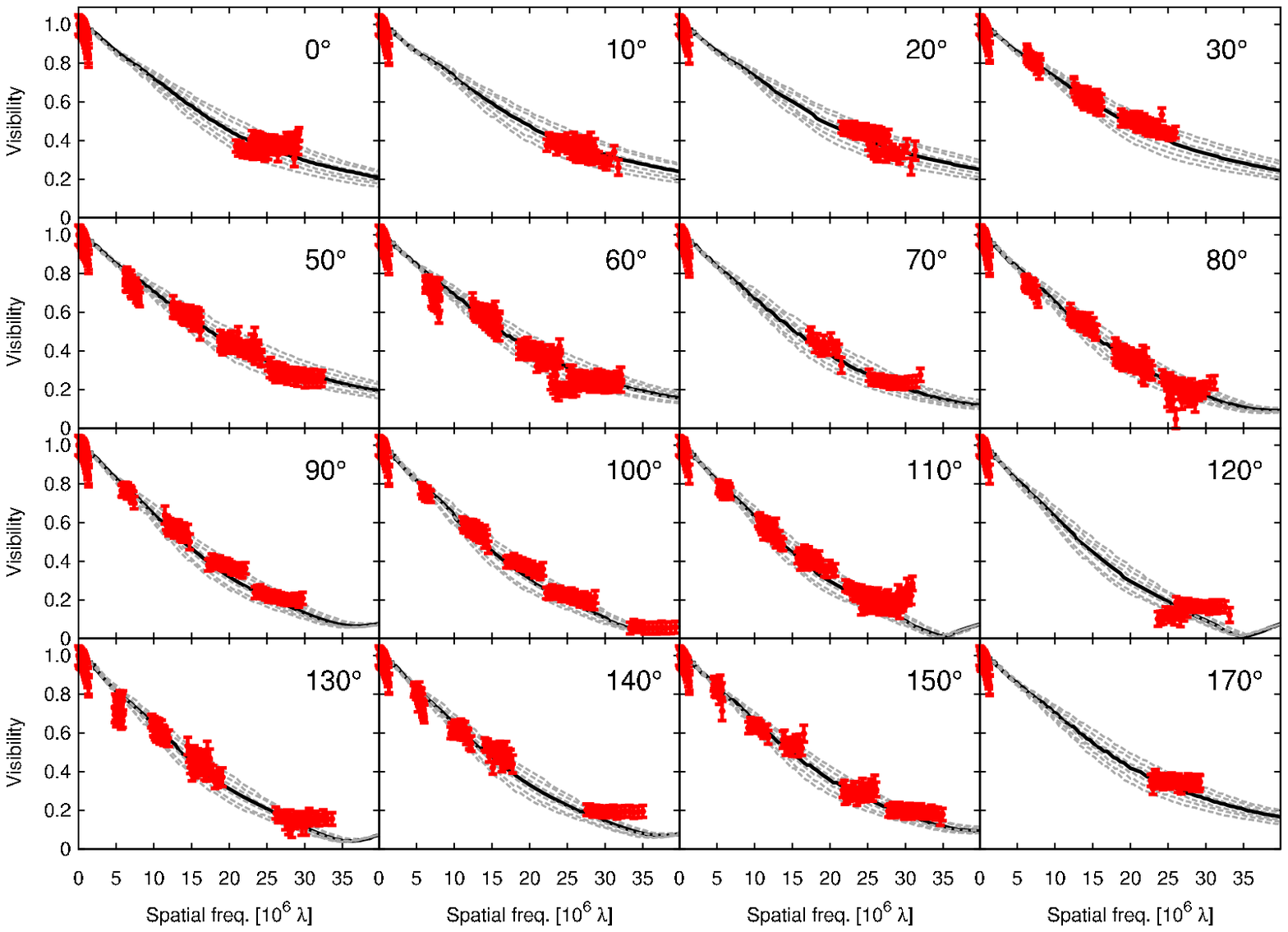}
  \caption{{\it Top:} Model image ({\it left}, $\lambda=2.25~\mu$m) and
    SED ({\it right}) corresponding to our 2-D 
    radiative transfer model.
    {\it Bottom:} Comparison of the measured 
     and the model visibilities for the wavelength range
     covered by our AMBER observations ($1.95$ to $2.55~\mu$m, grey dashed lines).
     The black lines show the visibility function for the central wavelength 
     of $2.25~\mu$m.
  }
  \label{fig:modelRTVIS}
\end{figure}

In order to confirm that the proposed scenario, 
including a high-mass young stellar object, a compact disk,
and an extended circumstellar envelope, can fit not only our
interferometric observations but is also consistent
with the measured SED, we performed detailed radiative transfer simulations.
To obtain an initial parameter set for our modeling 
procedure, we used the pre-computed grid of 
2-D radiative transfer models by Robitaille et {al.\cite{rob06}}.
The model parameters are then put into an adapted version of the
radiative code by Whitney et {al.\cite{whi03a}} in order to compute
model SEDs and synthetic images, from which we compute visibilities
and closure phases for comparison with our interferometric observations.
After some parameter adjustments we find a parameter set
which reproduces the SED and the measured
visibilities and closure phases reasonably well (Fig.~\ref{fig:modelRTVIS}).
The model includes a central star with a mass of $\sim 18$~M$_{\sun}$,
a circumstellar disk with a curved rim and an inner dust truncation
radius at 6.2~AU, and an extended envelope with bipolar outflow cavities.
For a detailed description of the model components and the best-fit parameters,
we refer the reader to our recent science {paper\cite{kra10}}.

\section{CONCLUSIONS}
\label{sec:conclusions}

Probing the dust distribution around a young high-mass star on the 
smallest physical scales yet, our study provides compelling evidence 
for the accretion-disk hypothesis in high-mass star formation 
and permits a first characterization of the geometry of the innermost disk regions.
Furthermore, our study demonstrates the benefits of a 
multi-wavelength (2-870~$\mu$m) and multi-scale imaging approach, 
which allows us to relate the detected AU-size compact disk with the
outflow signatures detected on parsec scales,
yielding a global picture of the environment around this massive young star.

With the available near-infrared interferometric instrumentation, 
the number of accessible high-mass YSO targets is rather limited, 
in particular due to the current sensitivity limits and the 
need for off-axis telescope guiding.
However, the number of accessible targets might increase considerably
with the upcoming generation of infrared interferometric instruments.
For instance, the implementation of infrared wavefront sensing units
would enable interferometric observations on a large number of targets 
without suitable visual guide stars.
For VLTI, this implementation of near-infrared wavefront sensors
($H$- or $K$-band) is planned in the course of the GRAVITY {project\cite{hip08}}.

Besides these improvements for telescope guiding and beam injection,
further progress could be achieved by observing in an optimal wavelength regime.
Near-infrared ($H$- and $K$-band) interferometric observations provide an excellent angular resolution 
and already enable phase-closure imaging observations,
but are restricted to a small sample of targets.
In the mid-infrared $N$-band (8-13~$\mu$m), the brightness limitations
are significantly relaxed, while at these wavelengths the disk contributions 
get more contaminated with emission from the envelope and 
the angular resolution becomes insufficient to study the inner disk truncation region.
Therefore, it is likely that the optimal observing wavelength for many young high-mass stars is
between the $K$- and $N$-band, for instance in the $L$- or $M$-band.
Two-telescope $L$-band interferometry was recently demonstrated at the
Keck Interferometer\cite{rag09} and will also become available with the 
VLTI second-generation instrument {MATISSE\cite{lop08}},
enabling four-telescope $L$- or $M$-band interferometric imaging on a large sample of massive YSOs.

\acknowledgments     

This work was performed in part under contract with the California Institute of Technology (Caltech) 
funded by NASA through the Sagan Fellowship Program. We would like to thank the ESO Paranal staff 
for their support during our observing runs and their enduring efforts in improving VLTI. 
This paper is based on observations made with ESO telescopes at the La Silla Paranal Observatory 
and used archival data obtained with the Spitzer Space Telescope, which is operated by the 
Jet Propulsion Laboratory, California Institute of Technology under a contract with NASA, 
and on data acquired with the Atacama Pathfinder Experiment (APEX). 
APEX is a collaboration between the Max-Planck-Institut f\"{u}r Radioastronomie, ESO, and the Onsala Space Observatory.


\newpage 

\bibliography{kraus}   
\bibliographystyle{spiebib}   

\end{document}